# A GLANCE INTO THE FUTURE OF HUMAN COMPUTER INTERACTIONS

Umer Farooq, M. Aqeel Iqbal and Sohail Nazir

Department of Software Engineering
Faculty of Engineering and IT, FUIEMS, Rawalpindi, Pakistan
umer_ms13@yahoo.com, maqeeliqbal@hotmail.com, sohailnazirdar@gmail.com

*ABSTRACT*

*Computers have a direct impact on our lives nowadays. Human's interaction with the computer has modified with the passage of time as improvement in technology occurred the better the human computer interaction became. Today we are facilitated by the operating system that has reduced all the complexity of hardware and we undergo our computation in a very convenient way irrespective of the process occurring at the hardware level. Though the human computer interaction has improved but it's not done yet. If we come to the future the computer's role in our lives would be a lot more rather our life would be of the artificial intelligence. In our future the biggest resource would be component of time and wasting time for a key board entry or a mouse input would be unbearable so the need would be of the computer interaction environment that along with the complexity reduction also minimizes the time wastage in the human computer interaction. Accordingly in our future the computation would also be increased it would not be a simple operating system limited to a computer it would be computers managing our entire life activities hence fall out of domain of present computers electronic based architecture .In this research paper we propose a model that shall be meeting the future human computer interaction needs possessing linguistic human computer interference environment based on surface technology, automation and photonic computing, which would be reliable, efficient and quicker satisfying all the future artificial intelligence pre requisites.*

## 1. INTRODUCTION

Effective from the root of human advent computers had been playing exception role in human lives. As the humans progressed the corresponding increase in computers efficiency was observed and so its induction in human lives increased. We had reached a stage that now computer are every where not as an exception thing but as an integral tool of our lives. The computer's induction evolved the working output of humans and so does its adaptation pace for the facilitation purpose was observed in all areas of life. We are currently at the verge of transformation of manual life to the automated life. Moreover on account of human facilitation the computer were inducted at various stages of human task but its induction was some how prioritized and the human computer interaction was kept aside [1]. Though the computer performance has boosted a lot and so its induction but objective of human facilitation tough worked upon but still not given the required focus as it worth to be.

Apparently the human computer interaction has evolved a lot from the initial form of computers but still wide gap is left that needs to be covered. The upcoming future will be the domain of artificial intelligence referring to the induction of computers even at the very basic human task. In the futuristic artificial intelligence environment the slice of compromise on friendly human computer interaction may lead to failure of the specific product on account of substandard human computer interaction.  Moreover beside providing facilitation to the common man the computer were meant to use by the common man. Presently induction is the main motive is the computer induction though an important one but it can not make us neglect the enhancement of user friendliness. Current development focuses on the induction of computers meant to me tackled by the ones having specialized skills in particular technological domain. Which tough provide the effort reduction and facilitation in undergoing tasks but unnecessarily makes a technological background vital to make use of the computerized machines. [2]

Presently the operating system offers us a user friendly environment and all we have to do is to select a task from the graphic user interference and execute it.  Apparently seeming to be a normal task may require thousands of if not in millions binary operations. Till the present state of computers induction in our lives it can be appear to be a bounty but imagining the same thing in the future it does not workout. When our entire life will be automated we can not afford to select a mouse then move it cursor and then click to execute that. As such by then would be a minute task of life so much exertion and time wastage for a minute task it won't worth it. Simultaneously the computation executed by the computers in the predicted artificial intelligence era would not be a printer command by a keyboard key or a mouse click it would be gesture interpreted one so it would be involving sub sequent heavy computations. [1][2]

## 2. AUTOMATION

Automation the backbone of intelligent machines has reduced the human efforts to a great extent. Still the reliability of automated systems is a major concern. It may be compromised while performing a normal routine task, but coming to critical tasks there exist no space for errors. In critical decision support system a slice of error could yield to be a fatal one [3]. So automation in such scenario requires extreme perfection. So what is the end shall we compromise on induction of automation in our machines or identify the drawbacks and correct them.

To attain full benefits of automation it shall be at its highest level but as the level of dependency increases the risk factor increases, the whole task to be undergone is left up to machine intelligence. Table 1 shows the machines independency in performing the task at different levels of automation.

Now if the machine is error free the task is undergone with perfection, but the other scenario in which an error or malfunction occurs it may lead to a havoc. Then if we are calculating a regular bill it may be compromised and the effect can be healed. On other hand the error occurred was in the air defence system the consequences for sure are fatal. Moreover in case of automation bias if it is accepted by human operator and the task undergoes without error detection the results could be even worst [3]. So an intelligent automated system design to reduce the human work load an error in such circumstances could even further facilitate the failure.

A way out is this that we compromise on automation level and minimize the chances of any error occurrence by human involvement. Eventually this will lead to a safe exit but in doing so the efficiency will be compromised. By induction of human in process the time resource is compromised as generally the human take more time than a machine would which too is another factor that cannot be afforded.

Table 1: Role of Operator at Different Levels of Automation. [3]

| Automation Level | Type | System Function | Operator Tasks |
|---|---|---|---|
| 1 | Manual | No automation support or information | Decide, act |
| 2 | Decision support | Automated system provides information but does not execute actions | View recommendation, decide, act |
| 3 | Consensual automation | Automated system highlights recommended choice; user makes any selection | Concur or select other option |
| 4 | Monitored automation | Automated system executes the choice. User has veto power for some fixed period of time, but system defaults to its choice if the user does not act. | Veto if nonconcur |
| 5 | Full automation | System acts and provides no veto authority to the user. | Observe full automation |

No matter up to what extent evolution in automation domain is done it is still domain what it is programmed for. A machine can be programmed in a manner that what is done is considered an intelligent it cannot be made completely independent [4]. While implementing design of an automated system prioritizing the benefits one must remain in contact with the limitation of automation.

The flaws of automation rather limitation with low level of automation may be a less threat but with the increasing dependency on automation may even prove to be fatal especially in case of defense related computations. General merits of automation are as under:-

## 2.1 Quick computation

The speed of computation done by computer is far ahead as compared to human capabilities. Hence a task done by an automated environment would be done in lesser duration as compared to duration taken by human to do the same.

## 2.2 Deductive reasoning

Computers if programmed to work accurately remain strict to it as they do not posses free will the way human do. Hence if all the computation done is as programmed by strict adherence to procedure program such scenarios reduced rather nullify error chances.

### 2.3 Multitask handling simultaneously

The feature of multi processing in computers enables multiple procedures to be processed simultaneously without any inference and malfunction in computing results.

### 2.4 Repetitive and routine tasks

Computers if meant to process a repeated task need to be programmed only once and they keep repeating the task. All repeating procedures are undergone in a far better way a human would have done so with a leap in speed, efficiency and perfection.[3], [5]

Apparently seeming to be bounty automation has a lot of factor that if not managed may spoil all the taste of its benefits. Following are considered to be demerits of automation:-

Technology pace never pauses, there is no mile stone that could be regarded as the limit of technological growth. Hence there is always an enhancement in all of the technological products especially computers. Automation being directly related to technological growth so requires frequent updating to remain functional as remain outdated is not acceptable. Due to mentioned scenario the cost of automation is never known.

### 3. SURFACE TECHNOLOGY

Surface technology refers the merge of physical and the virtual world interacted by a vision as a mode of interface. Surface technology means human computer interaction by elimination of input output devices that is a display with a touch sensitive feature playing role of input output devices. Talking about the surface technology the user is meant to deal directly with the objects by eliminating all the graphic user interface mediums such as icon, windows e.t.c with the objective of reduction of gap between the physical and the virtual world. By the virtue surface technology a user interacts with digitalized world by just a finger touch. [3]

Surface technology can be classified into two generalized components one concerned with the display and the other one with the interpretation of user signals via a touch sensitive mechanism. The component concerned with the display can be built on display platform such as front projection or the flat panel display. On contrary the user signal interpretation component is based on image taken by image sensing cameras generally the infrared based. In order to do so cameras are adjusted in a manner that whole screen is covered by the cameras. [4]

Infrared based sensing can be done in three generalized ways that include diffuse illumination, frustrated total internal reflection and diffused surface illumination. [13]

Diffuse illumination could be utilized by both rear and front illumination. In the case of rear illumination it works on the principle that when an object touches a surface it alters the reflected light which is being reflected by diffuser this change in reflection is then sensed, interpreted and then computed. On contrary front diffused illumination infrared is projected on screen and as the object touches the surface a corresponding shadow is casted which is interpreted further for computation of result. Figure 1 describes the placement of different components of surface tecnology based on diffused illumination.

International Journal of Computer Science, Engineering and Applications (IJCSEA) Vol.1, No.3, June 2011

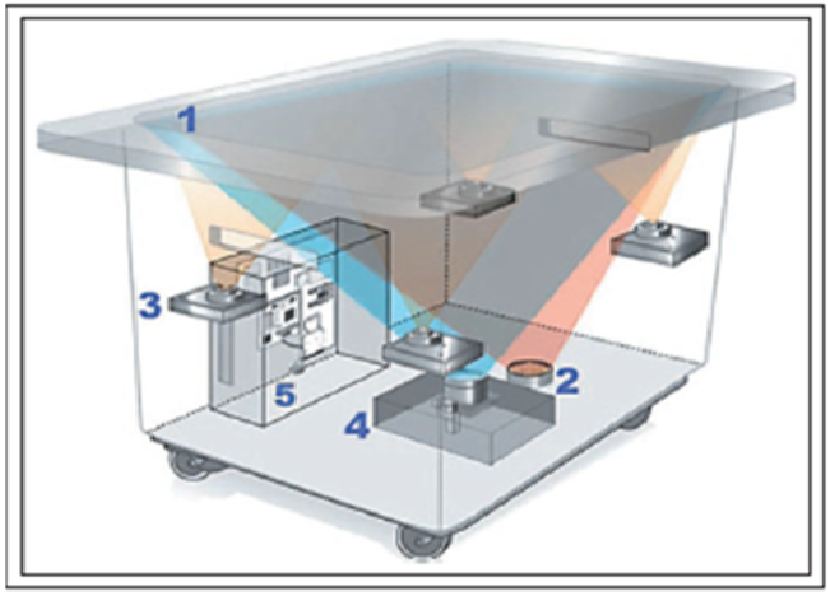

Figure 1: Working Of Diffused Illumination Based Surface. [3][4]

Frustrated total internal reflection is based on the reflection principle of Snell's law Popularized in the touch-screen. Based on the difference in reflective indexes of different materials portion where an object touched a screen at that point the total internal reflection does not occurs and the scattered light is identified by infrared based camera imaging. Figure 2 shows the working of frustraed total internal reflection.

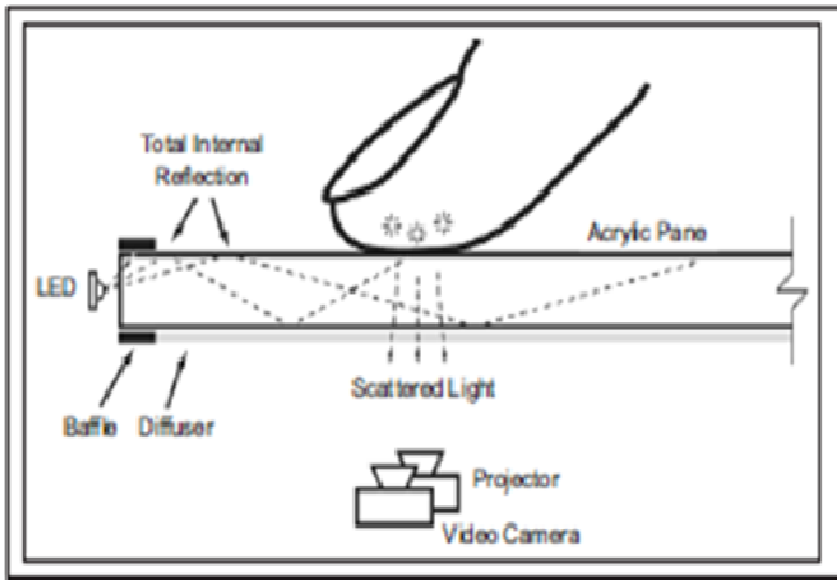

Figure 2: Behavior of Light at the Touch Spot in Frustrated Total Internal Reflection. [3][4]

Diffused surface Illumination is composed of acryl glass used for the even distribution of the infrared light across the screen. The acryl glass is composed of tiny particles, each of them performing as a mirror that evenly distributes the lights. As the user makes a contact with the screen the light gets scattered this is observed by camera and accordingly interpreted for computation purposes.

Though not launched commercially at the present but the breakthrough that would be brought by the commercialization of this technology is evident from the launch of apple I phone which is based on vision based technology. It is expected that a substitute for the input mouse and key board interaction with the standard graphic user interface for sure will enjoy a vast market boost as is evident from the market statistics of I phone that defined new dimensions in the domain of phone technology. [4], [5]

Apart from the user convenience benefits there is another characteristic of the surface technology that it could be built out of less expensive components and simultaneously open source software support for it indicates the development potentials in this technology [4], [5]. Currently there are four generalized feature offered in surface technology as offered by Microsoft which is regarded among the benchmark setter of the surface technology that include direct interaction, multi touch reorganization, multi user support and object reorganization.

Direct interaction offers user communication directly with the screen eliminating the role of input devices such as mouse or a keyboard. This feature offers multiple processing points simultaneously unlikely a mouse that offer one point where the cursor is to be processed a multiple contact points are available facilitating certain user to interact at a common time slice. Object reorganization offers detection of electronic devices such as mobile phones and digital cameras that are tagged. User can interact with the data in the tagged devices with the features offered in the surface technology.

Surface technology is meant to bring revolution the interaction of humans and computers. Glancing future it may be implemented in the commercial appliances, vehicles, homes, offices, shopping plazas, hotel reception. Hence every where it will be applied it will prove its worth and intensively support human in caring out there daily routine activities in far efficient way current technological system offer. Eventually a point may come when all the persons in the mentioned places may be replaced by the intelligent surface technology screens and that will for sure be the predicted upcoming era of artificial intelligence.

### 3.1 IMMO Touch

Imagine going to a real estate office the maximum details one can get of the desired property is a detailed map and in rare cases a picture of the property [6]. If one get all this service one get pleased and rate the service satisfaction level to be ultimate.

Product of surface computing Immo touch if inducted in real estate office for getting the desired information could reveal wonders. It shall be offering advanced searched information that too with an interactive search medium. Giving input by a finger touch and getting desired results appeared on table top surface. The results wont be in form of a map on a per for which one has to first hold it then adjust it to a suitable positions to be read rather on a table top surface and for zooming one wont be putting strain on eyes rather but by using fingers to zoom in or out. When done with the search one can undergo desires filtration and removal of unnecessary information and eventually instead of memorizing the induced results email it to your email address. [6]

Other offered services include detail review of property by experts, geographical location views by the help of video streaming on the surface. The table top Immo touch also supports transfer of files directly from surface to USB. Figure 3 shows a user interacting with IMMO touch.

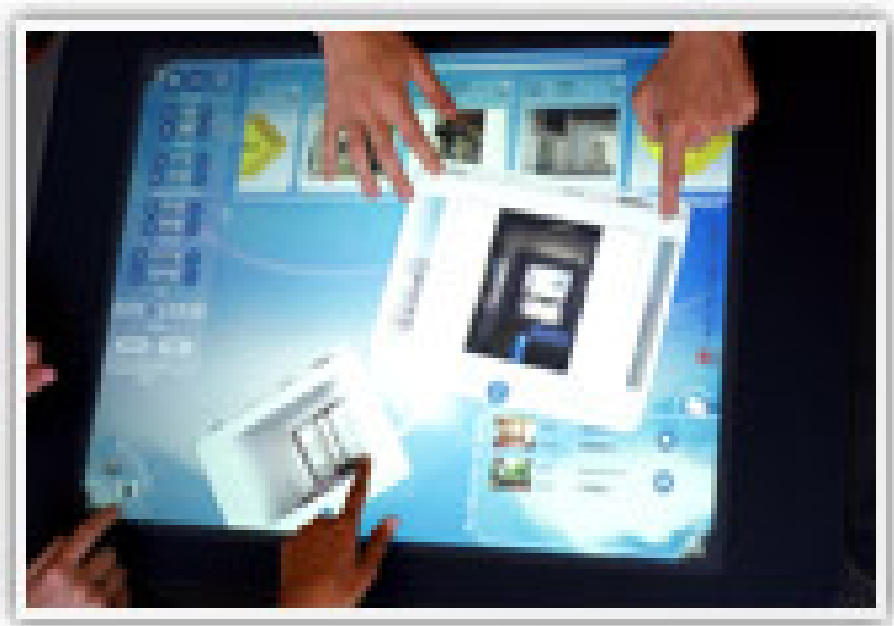

Figure 3: Users Interacting With Immo Touch. [6]

If ones intend to undergo the comparative analysis of the presently manual services offered and the one offered by the surface technology products the quality of surface technology speaks for itself. The surface technology induction in real estate marketing sector shall revolutionize the service and user friendliness offered rather by present systems.

## 3.2 RESTO touch

Getting into restaurant having available staff to assist you is ordering the desired meal is what the best manual life is presently offering. In undergoing the selection of desired meals we have to read the menu and eventually order the desired food. In the process we had to interact manually with the menu then make search for the desired items which is eliminated by the RESTO touch. Figure 4 shows the RESTO touch features. [7]

Resto touch eliminates the interaction with the waiter to get the menu and then the searching for the desired choice. When a customers come they just sits on their place view all the menu items and order directly from their table surface in a very interactive and convenient way by just finger touch. Aside it provides pictures and videos of the available meals that make better choice as are provided with detailed information of their choice. Features offered for wine lovers along with enabling them to order directly from table it also provides detailed information of their choice and available brands. It also shows history, pictures and videos of all the available stock in order to assist user in making appropriate choice. [7]

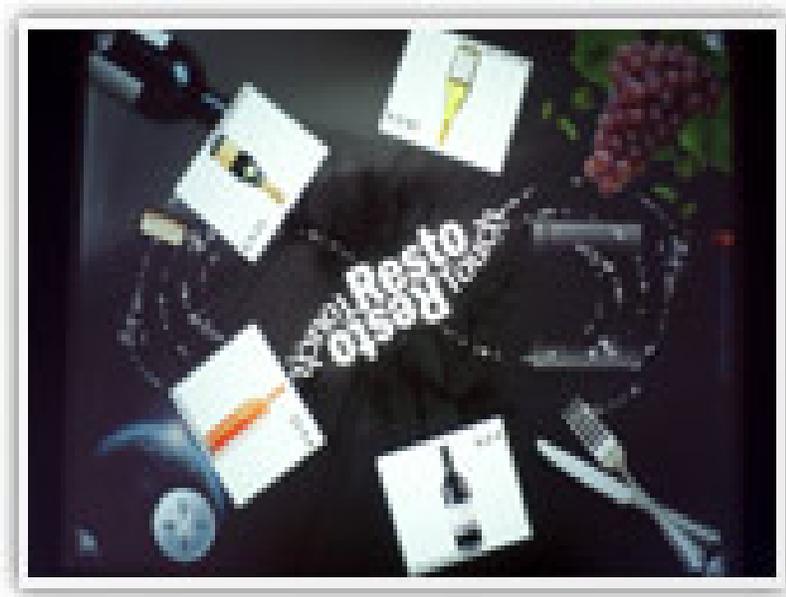

Figure 4: Users Interacting With RESTO Touch. [7]

Tourist guide feature is also available for the customer so that meanwhile enjoying food and drinks could also gather the required information for travelling nearby areas. The information is displayed in 2d and 3d environment that offers better understanding for the tourists. Option for future booking is also available that is compiled in real time and implemented directly on database.

## 4. ARTIFICIAL INTELLIGENCE

Artificial intelligence or the world of computers domination is the forecasted computer world in which computers are meant to be intelligent. In the artificial intelligence computer will breach the restrictions of performing within the prescribed feed program and instruction and would be deducing and then executing appropriate action for a certain event [8]. The computer shall be performing all the tasks of our daily routine with utmost independence eliminating the human induction in the tasks.

Apparently appearing as a blessing of technology and a heavenly lifestyle by induction of artificial machine still a lot of issues are yet to be overcome in order to make this dream into reality. Generally the intelligence in a computing machine is evaluated by comparing its computation efficiency and approach with respect to human approach towards executing a response to an observation. The issue arises how a computer can interpret perceived data as a human can. Contrasting human brain thinking capabilities with the computers computation certain issues arise. A human initially observes then after deducing the appropriate result for the observed event and required execution is undergone which is referred as human intelligence. Conversely if a computer is observing an event all it does id that initially interpret the observation and then perform the relevant action if it is programmed for that event. If a situation occurs that the computer was not programmed for then it remains a dummy as the event is unknown for it. [8][11]

Human respond to an action based on reasoning and on spot decision not following a set of instructions. On other end the computers respond exclusively as per set of instruction they are programmed for. Moreover a human's response may be different for a same event like if one is testing an air surveillance system an aircraft detected may be ignored by a human operator. The same event if encountered by radar is regarded as a threat the way any enemy aircraft is considered without distinguishing it to be a friendly one.

The issue arises how come we can bring computer's computational power to an extent that it can execute an observed event like a human brain does. For sure this is a hectic and even this is impossible to make a computer recognize all the events to it and the biggest constrain is that of the memory among the memory constrain there are two aspects the huge amount of memory that shall be required and the other is the way it would be upgraded. A human brain is updated or one may say developed by sequence of events it encounters. The event is stored in hierarchy which is developed over a period of time and developed from event to event.

Assume a simple scenario of a man holding a gun. In such case a human may interpret the one as a security personal or a miscreant based on deduction of observed behavior and body language of the person. If a similar case is encountered by a robot it would deduce result by constrained by the set of programming instructions it is programmed. If programmed that a man holding a gun is miscreant it would be a miscreant and if man with a gun programmed to be a security personal it would interpret it to be a security personal. Similarly if we have an automated gun that locates a target and fires, now if we command it to fire I will locate any human there and fire at him irrespective of him being friendly or an enemy as it can't differentiate between them. Conversely if a human army is assigned the same task it will target only the enemy when the same order of fire is given to them. From the mentioned scenario one may say that it's the decision power that lacks in the computing machines. The decision power is the main concerning issue in and considered as the main area that need to be worked on for the real implementation of artificial intelligence in the real world. [10][11]

Moreover the decision of execution of an event can not be programmed at once. This process is developed and enhanced over a period of time in which the events occur and there responses based on cognition process are defined. Now if we intend making computer to think like a human we need it to respond to actions like a human do in short compute like a human brain does which refers that for achieving which computer programming to make them intelligent shall be requiring periodic updating.

The achievement of artificial intelligence shall be requiring expertise from all the disciplines of engineering, medicine, computer science e.t.c. when all the expertise from the mentioned fields when integrated altogether will form a single state of the art product that shall be initialization of artificial intelligence. The way it requires expertise from all the fields of life the same will be its impact on our lives and our living standards above all will be redefining the way we interact with the present miracles of the technology. Artificial intelligence shall be extending the applications of the technological miracles to new limits rather redefining there as a net result yielding an ever friendly interaction between the user and the technological products. Chart 1 shows the parent disciplines of artificial intelligence and its domains.

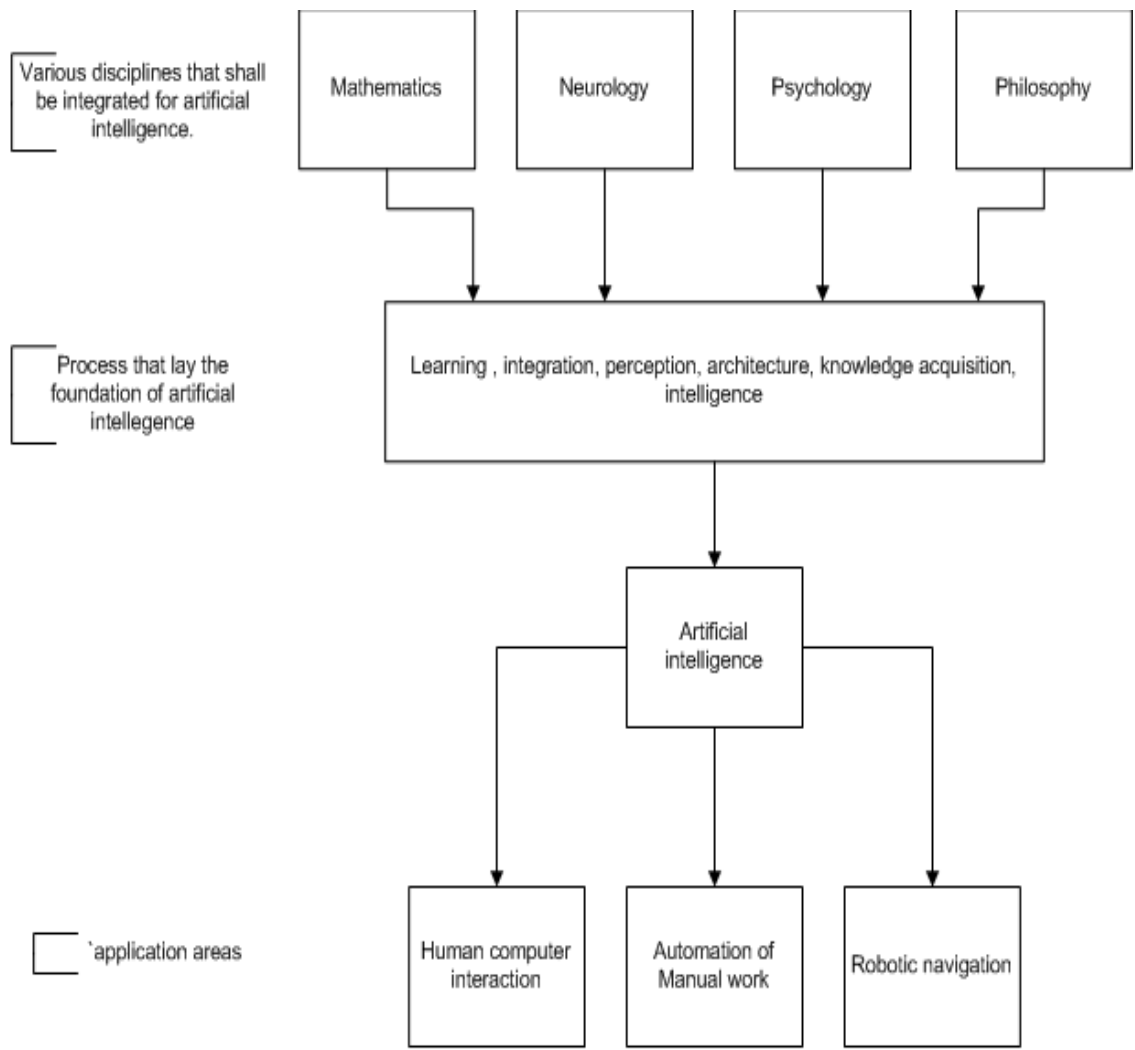

Chart-1: The Chart Shows The Parant Diciplines Subjects Covered In Artificial Intelligence And The Application Areas Of Artificial Intelligence. [8][11]

Implementation of artificial intelligence requires immense amount of data storage along with a marvellous speed of computing. The present electronic architecture appears to be at verge of its max performance potential. We had sqeezed the micro processor architecture to a limit that no more shrinking will be practicaaly possible in the near future. Figure ML shows the Moore's Law prediction regarding the number of chip per unit space. Same scenrio is observed in the memory domain as the expected amount of data that is needed to be programmed in the futuristic artificial intelligence devices would be beyound the capacity of electronic based storage. So the need of the hour is an alternate for electronic based computing and memory architecture. Figure 5 shows moores law predictions about chip density [14]. An alternate to the electronic based computing and memory storage is an photonic based architecture that along with offering high speed computing also enables far more amount of data that an electronic nased architecture would have.

31

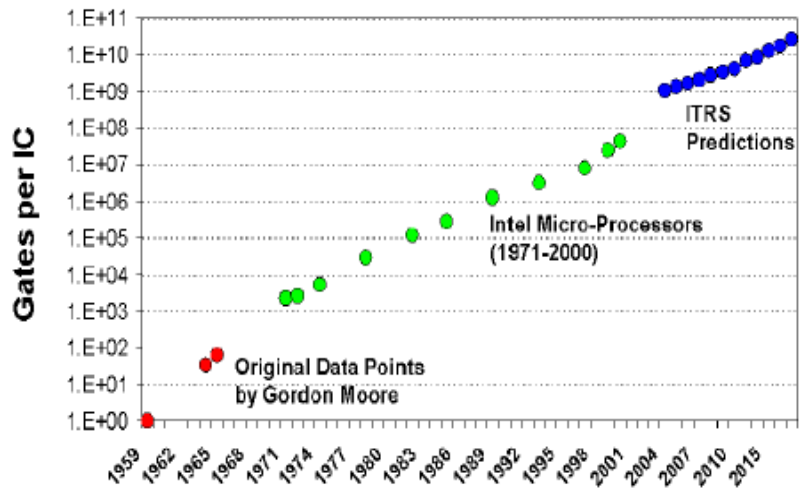

Figure 5: Shows Moore's Law Prediction Regarding Chip Density.[14]

## 5. EMBEDDED COMPUTING

As the technology evolve the computational products kept on modifying and new architectures kept on emerging [9], [12]. Progress in the technology kept on yielding state of the art products and computers became generally available to majority of users. As the technology progressed computers initiating taking over all the tasks that human once used to perform manually.

The progress in computing led to microprocessor architecture that minimized the size of computational device. Eventually they become portable and embedded computing was born. Embedded computing is generally referred as portable computing devices possessing relatively low processing power, limited memory and precise domain of applications. But the present embedded systems had nullified this generalize concept regarding embedded computational devices. Now the embedded computing based devices are performing state of the art functionality with dazzling computational power and enjoying a vast applications domain.

The advancement of technology has made the quality standard with high level of reliability. Now the systems are becoming more and more complex in architecture and performing ever more complex tasks. The application of embedded computing includes all fields for sake of example automotive industry the ignition system, engine control e.t.c. customer electronics automobiles, cameras, sensors e.t.c. Embedded computing has also yielded miracles in the medical field and automation domains.

As per the present state of development may we assume that we had achieved the state of astounding limit? But forecasting the future requirements it is not even worth considering. The upcoming era will be of the artificial intelligence in which the computing wont is limited to a particular task. It would be state of the art computing devices that would be doing all the tasks humans are doing manually at present. So the new era shall be requiring new computational techniques that shall unable us to enjoy the bounties of futuristic artificial intelligence era. Chart 2 explains the designing and development lifecycle of embedded systems.

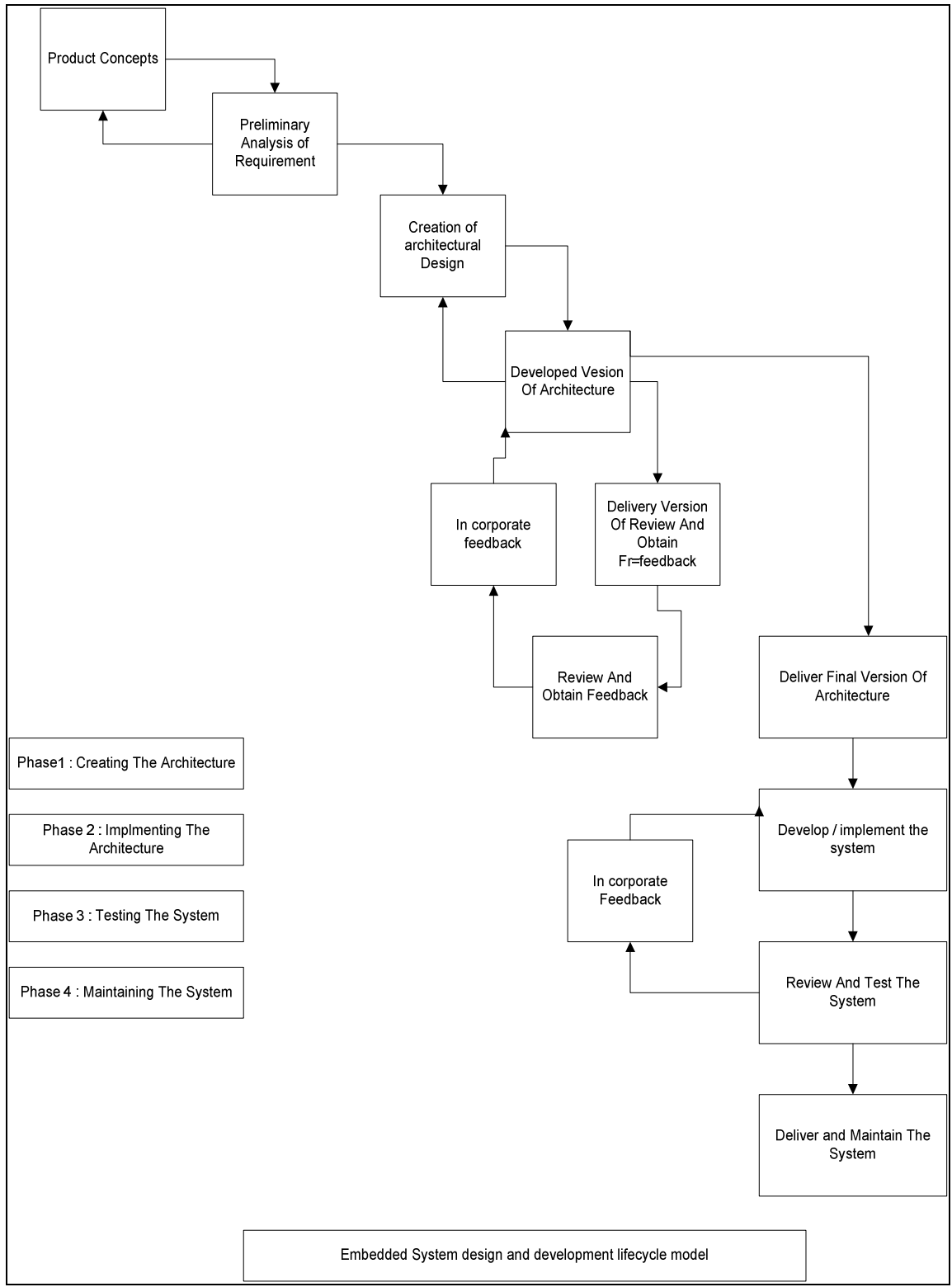

Embedded System design and development lifecycle model

## 6. PHOTONIC COMPUTING

Analyzing the required enhancement in the upcoming era apart from traditional input devices based human computer interaction a revolution is also required in the field of computer working power. Moreover the machines are required to be made intelligent in a way that that maximum of the task is automated and that's only possible by induction of new state of the art computational logics. In other terms one may conclude that the future doest demand the computer that an engineer or a specialized person could operate. The role of engineering and technical people shall be restricted till development and afterwards it shall be in a form that it be handled by the general public users. So what the way to such development the only way it seems functional is that the computing products shall be built in such a manner that it itself become an advisory to user. The only way this requirement seems to be satisfied is that the computer shall be built to provide interface very close to human nature. [14]

Linguistic computing can be a solution that does not require user to be specifically trained or even if required just a general brief one to manage the computational device. Currently we interact with mouse that once appeared to be a milestone in redefining the human computer interaction. By then till present it modified a lot from cord to cordless and with even a smoother touch but still it not of worth to compete the upcoming era. [3] [14]

Virtual computing an implementation of linguistic is appearing as a bridge of transformation of manual human task to a fully automated span of artificial intelligence. Microsoft has launched its surface technology products and appeared as the pioneer in redefining the human computer interaction.

Yielding a state of the art architecture giving a user friendly interface to the end user may we would satisfy the user requirement but in doing so the required computational operation shall boost enormously. Focusing on the current products it could be accomplished by using electronic based computational logics but can we implement the electronic based computational for the future products. But will the future systems based on linguistic computing will they be of the same nature. Surely even saying such a statement sounds stupid. The future computation shall be encountering massive amount of data that too with required parallel processing, individually computed without even a bit of error accommodation. Will the electronic based computation allowing us to do so certainly not.

One way is enhancement of the electronic based computation by multi tasking, multi processing and enhancement of processor power but how could we achieve it. Eventually we will be facing the electrons barrier achieving which no further enhancement will be possible. The crystal clear is to stop dreaming it's just not an electrons worth to full fill such requirements. Surely electrons cannot be extended as we have forced the electron to its maximum potential. Electron based computing being facilitating our lives to great extent but as a nature of human there is no limit or a restriction to human development pace. [13] [14]

As in present we have reached the stage of nano technology now can we go further sure we can but will the electron allow us? It's certainly not permitted they more we push electrons together the more unreliable our computing becomes. As the electrons tend to react with each other resulting is distortion of data that for sure is unbearable. [14] Figure 6 shows optical communication in an optical computer.

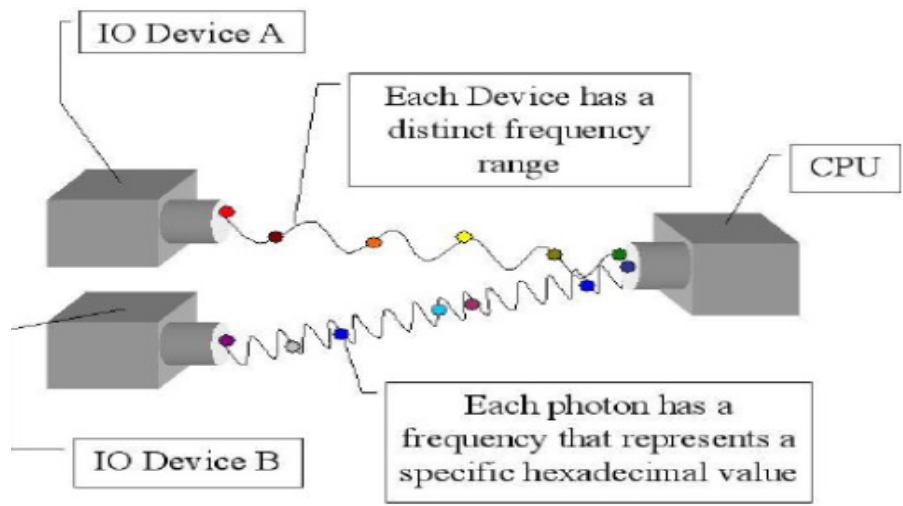

Figure 6: Optical Communications between Input Output Devices of an Optical Computer, [14]

Photon the fundamental particle of light appears as an alternate to electron. The flaws rather the short coming of electrons can be addressed in photonic based computing. Figure 7 shows components of an optical computer. Some general comparison of certain features of electronic and photonic based computing is as under:-

1. A single photon can be represents in 16 different and distinct ways based on difference in wave length. Implementing photons we can address a large amount of data corresponding to electronic based computing. If we use a 32 bit system by using electronic computing we can address 2^32 bytes per unit time on contrary by doing the same with photon it could be as much as 16^32 bytes per unit time which it self speak the worth of photon with respect to electron.

2. Photon can move from one point to another without need to apply potential for movement, in a straight line with the speed of light. On other end electron require potential to move from one point to another, that too in circular orbits furthermore with the comparatively sluggish speed of electron.

3. In the domain of computer interconnection pathways commonly refer as busses if we use electrons the max we can get is traditional 64 or 128 KB bandwidth busses. Conversely if we come to photonic interconnections we get 35 billion separate non interacting pathways which for sure can't be even assumed to be ever being achieved using electronic interconnection.

4. Experimentally it is proved that photons can transmit up to 200 GBs of data per second while it seems a fantasy if we even think of comparing electron as a competitor to photon in data transfer rate.

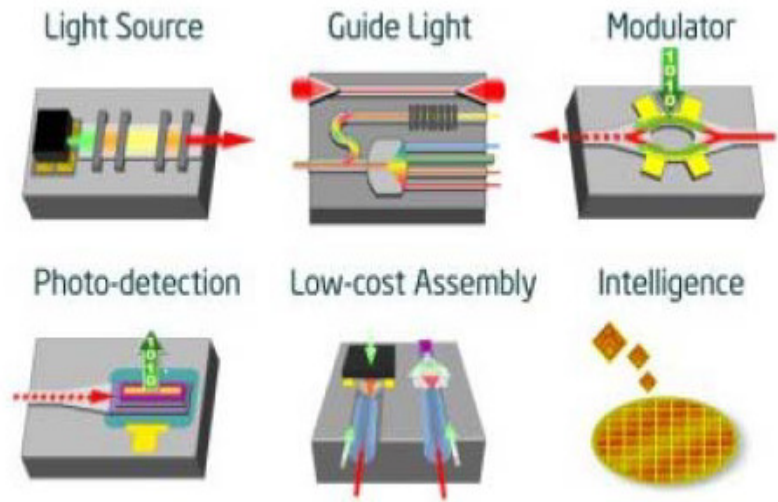

Figure 7: Main components of an optical computer. [14]

One may say that tough we had pushed the limits of electron it's just the verge of extinction of electrons from the computation. It's a matter of fact that it's the end of electrons but shall our computing based on electrons shall it be a greatest ever milestone achieved. Surely it's foolish to even think of, the upcoming time will be the artificial intelligence rule the era of machine. And the revolution will be brought by photonic computing that surely could be called the next generation of computing.

## 7. CONCLUSION

Future of computing defines the glance of the initial interaction between human and computer and its co relation with the present technological miracles. It is assumed that we are in the technological era where we are enjoying the best technology could offer. But wait are we sure that this is the best technology could offer. Tough we had offered user with a high level of user friendliness but a lot more needs to be done. Surface computing is a step from the traditional computing to the fanatic world of artificial intelligence. At initial steps the electronic based computation appears to be supporting but the future high speed computation speed and huge memory requirements appear to be beyond the scope of electronic based computing architecture for which photonic computing is an proposed alternate.

International Journal of Computer Science, Engineering and Applications (IJCSEA) Vol.1, No.3, June 2011

**AUTHORS PROFILES**

**Umer Farooq**

Mr. Umer Farooq Is Student Of BCSE Program Of The Department Of Software Engineering, Faculty Of Engineering And Information Technology, Foundation University, Institute Of Engineering And Management Sciences, Rawalpindi, Pakistan.

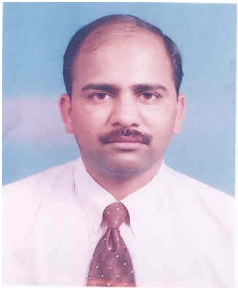

**M. Aqeel Iqbal**

Mr. M. Aqeel Iqbal Is An Assistant Professor In The Department Of Software Engineering, Faculty Of Engineering And Information Technology, Foundation University, Institute Of Engineering And Management Sciences, Rawalpindi, Pakistan. He did his MS In Computer Software Engineering From National University Of Sciences And Technology (NUST), Pakistan And Has Many Research Publications In Well Known International Journals And Conferences.

**Sohail Nazir**

Mr. Sohail Nazir Is Students Of BCSE Program Of The Department Of Software Engineering, Faculty Of Engineering And Information Technology, Foundation University, Institute Of Engineering And Management Sciences, Rawalpindi, Pakistan.